# A Two-Step Deep Learning Method for 3DCT-2DUS Kidney Registration During Breathing

Chi Yanling[1*#], Xu Yuyu[2#], Liu Huiying[1], Wu Xiaoxiang[2], Liu Zhiqiang[2], Mao Jiawei[3], Xu Guibin[2*], Huang Weimin[1*]

[1]Institute for Infocomm Research (I2R), Agency for Science, Technology and Research (A*STAR), 1 Fusionopolis Way #21-01 Connexis South, Singapore 138632, Republic of Singapore {chiyl, liuhy, wmhuang} @i2r.a-star.edu.sg
[2]Key Laboratory of Biological Targeting Diagnosis, Therapy and Rehabilitation of Guangdong Higher Education Institutes, The Fifth Affiliated Hospital of Guangzhou Medical University, 510700, People's Republic of China, gyxyy@foxmail.com, 348215424@qq.com, {zhiqiang_liu2012, gyxgb}@163.com
[3]Creative Medtech Solutions Pte Ltd, Republic of Singapore mao.jiawei@ultrastmedtech.com
* Correspondence authors, # Equal contribution authors

## ABSTRACT

This work proposed KidneyRegNet, a novel deep registration pipeline for 3D CT and 2D U/S kidney scans of free breathing, which comprises a feature network, and a 3D-2D CNN-based registration network. The feature network has handcrafted texture feature layers to reduce the semantic gap. The registration network is an encoder-decoder structure with loss of feature-image-motion (FIM), which enables hierarchical regression at decoder layers and avoids multiple network concatenation. It was first pretrained with a retrospective dataset cum training data generation strategy and then adapted to specific patient data under unsupervised one-cycle transfer learning in onsite applications. The experiment was performed on 132 U/S sequences, 39 multiple-phase CT and 210 public single-phase CT images, and 25 pairs of CT and U/S sequences. This resulted in a mean contour distance (MCD) of 0.94 mm between kidneys on CT and U/S images and MCD of 1.15 mm on CT and reference CT images. Datasets with small transformations resulted in MCDs of 0.82 and 1.02 mm, respectively. Large transformations resulted in MCDs of 1.10 and 1.28 mm, respectively. This work addressed difficulties in 3DCT-2DUS kidney registration during free breathing via novel network structures and training strategies.

## Introduction

Medical image registration is a process that aligns one image to another originating from the same or different modality. This aligned image contains more spatial-temporal information, which is important for applications such as image guided surgery[1], disease monitoring[2] and risk prediction[3]. Registration between images of the same modality is mono-modal registration, and registration between images of different modalities is multimodal registration. Different imaging techniques are sensitive to different tissues in the body. Therefore, images of different modalities need to be registered with each other to provide complementary information. However, this is difficult because of the complex relationship between the intensities of the corresponding structures in the two images. Ultrasound (U/S) images are especially challenging due to their large motion, small field of view and scan quality. Nonetheless, 3D-2D registration is needed. The potential of deep learning on those issues has not been fully reached[4]. In this work, we proposed a two-step deep learning method to address 3D computed tomography (CT) to 2D ultrasound (3DCT-2DUS) kidney registration.

State-of-the-art (SOTA) methods[5] can be classified as supervised, weakly supervised and unsupervised registration, according to learning strategy or convolutional neural network (CNN)-based, deep adversarial network-based, and transformer-based image registration, according to baseline network architecture. The supervised registration [6] is trained to predict the transformation by using images and their ground truth transformations. Weakly supervised registration[7,8, 9] uses overlapping segmentations of anatomical structures as a loss function, which lowers the limitations associated with ground truth data. Unsupervised registration[10, 11, 12, 13, 14, 15] is trained by minimising a dissimilarity measure given a set of images and does not need ground truth transformations. CNN-based image registration[16, 17] trains a designed CNN architecture and learns the mapping between the input images and the deformation fields. Deep adversarial image registration[18, 19] consists of a generator network and a discriminator network. The generator network is trained to generate transformations and the discriminator network learns similarity metric to ensure that the generated transformations are realistic, or the input images are well registered. Vision Transformer (ViT)-based registration[20, 21, 22, 23, 24] learns the inherent relationships among data through the attention mechanism. Our solution is CNN-based unsupervised registration. We refer to registration as unsupervised learning because the registration subnet is under unsupervised training. The feature subnets are trained separately and not specifically for the registration task. They are independent feature extractors, and universal features are also applicable to our solution.

Most 3D-2D registration is supervised projective registration. The 2D image is the projection of the 3D volume. Miao[6] proposed using CNN regression to register 2D X-ray images with 3D digitally reconstructed radiograph (DRR) images. Ground truth transformations were available for their application. Foote[25] proposed tracking tumours using a single fluoroscopic projection via supervised learning-based method. The method densely sampled the CT volumes and calculated DRR projections with a linear combination of motion components via DenseNet [26]. Salehi [27] estimated the pose of 2D MR plane within the MR volume using supervised regression CNN. Liao[28] and Krebs[29] proposed employing reinforcement learning to conduct 2D/3D registration by learning a series of actions. Our method is sliced 3D-2D registration. No assumption of projective geometry is made, and no ground truth transformation is used for training. In addition, it faces the challenges of a very large potential search space. We address those difficulties using the proposed novel solution. Guo[30] proposed aligning 2D TRUS frame with 3D reconstructed TRUS volume using deep registration network. Their method used CNN regression to estimate transformation parameters and compared them with ground truth transformation for mean square error loss. It sampled a 2D slice using estimated transformation and compared the slice with 2D TRUS for auxiliary image similarity loss. The author found that unsupervised learning cannot result in stable training. Therefore, the network was trained on mono-modal images under supervised learning with the combined loss. Wei [31] proposed registering vessel labels on 2D U/S images to vessel labels on 3DCT/MR liver images using deep registration, which was a mono-modal approach. They labelled the image manually and did not analyse the complex relationship between the intensities of corresponding structures in multimodality images. The registration model was trained under supervised learning and followed by a conventional plane fitting process.

Deep affine registration uses an encoder structure[14], Siamese encoder structure[32] or ViT-based structure[33]. They are 3D-3D single modality image registrations and cannot be the full solution for 3DCT-2DUS registration. Moreover, for registration at multiple scales, two or three encoders are stacked together due to capacity limitations. In contrast, we propose using an encoder-decoder structure, which enables multiple-scale registration by hierarchical regression of transformation parameters from decoder layers. It allows more scales and benefits large transformation registration and fast model convergence.

Balakrishnan[10] proposed VoxelMorph to perform unsupervised registration. We adopted their baseline architecture and improved it to a hierarchical architecture to generate transformation parameters at multiple scales. Hu[8] proposed using image features to guide registration. Our work is conceptually superior to their work[8] in several aspects. The first is on architecture. In their work, original images are used to train the registration network, and the ground truth image segmentation labels are used to calculate the loss function. Our work uses a network to represent images and inputs image representations to the registration network. Both voxel-level image content and high-level features contribute to registration. The second is the loss function. A modality-independent neighbourhood descriptor (MIND)[34] is used to measure the image similarity of the CT-US pair. Here, we assume that a high-level feature can drive alignment close to its optimum and that MIND loss is continuous locally. A gradient loss is designed to regularise respiration motion smoothness on windowed U/S scans. The third is that they address 3DCT-3DTRUS prostate registration. We address 3DCT-2DUS kidney using a deep rigid registration. We pretrain a model by using an unsupervised learning cum data generation scheme and refine the model by one-cycle transfer learning. Heinrich[35] proposed a discrete 3DCT-CT registration, which used two steps of optimisations, unsupervised learning for global searching and a correlation layer for local optimal search. Our method integrates features, images, and motion metrics into the loss function and conducts one-step transformation estimation.

In this work, we contributed a novel deep learning pipeline for sliced 3DCT-2DUS kidney registration. It overcame two main challenges: registering images of different dimensions and imaging modalities. To address the dimension difference, the U/S images were first expanded to the same dimension of CT by zero padding. Because there were few spatial constraints between CT volume and 2D U/S slices compared with 3D-3D volume registration, it was necessary to move 3D CT effectively. We proposed using a rigid encoder-decoder registration network and hierarchical regression of transformations from each decoder level. Transformations for images from low to high resolutions were combined at the highest resolution via weighted translation and rotation. In addition to hierarchical regression, we designed a combined loss to drive image sequence alignment accurately via global deep kidney features and local modality-independent image features and smoothly via the transformation of consecutive frames. Moreover, to further improve the registration performance and ensure efficiency in clinical applications, we proposed unsupervised training the registration network in two steps: pretraining the model using the general training datasets and adaptively training the model for two epochs using specific patient training datasets by one-cycle transfer learning. Training data generation was proposed to generate image pairs for general training. To address the different imaging modalities, we proposed extracting deep kidney features on CT and U/S images for overall comparison and extracting modality independent image features for comparison with local details. The feature network was designed with handcrafted texture layers to reduce the semantic gap. Furthermore, we applied a time window on the U/S sequence to improve kidney observation on noisy images by including respiration motion information. Generally, the methodology addressed all issues in kidney registration during free breathing. To the best of our knowledge, this is the first deep learning pipeline for sliced 3DCT-2DUS kidney registration.

## Methods

The human kidney seldom deforms due to patient posture changes and respiration according to clinicians. Therefore, we used rigid transform in the kidney image registration. The proposed solution consists of 3DULBNet[36] and a 3D-2D hierarchical registration network. 3DULBNet was trained on CT and windowed U/S images on binary segmentation tasks separately offline. They were connected with the registration network to predict the CT plane that best aligned with the U/S images.

### Feature Network

The ULBNet is a 5-level U-Net with a residual block replacing the original convolutional layer (Appendix A). A local binary convolution (LBC) layer [37] was added to skip connections. The dropout rate was set to 0.2. For CT images, the patch size is 160×160×80, and the batch size was 1. The optimiser was Adam. The loss function was the negative Dice coefficient. The output layer was a convolutional layer with sigmoid activation, and its output was a kidney feature map of CT images. The feature map was a probability map of a pixel/voxel being kidney. It described global shape characteristic of the kidney. ULBNet[36] can be referenced for details of the method used to process CT images.

U/S images are noisy, and it is difficult to delineate kidneys from a single image. Kidney motion information is useful. Since the U/S sequence scans the kidney in respiration, the kidney motion cannot be observed in one frame but can be observed in a few consecutive frames. The U/S image sequences are experimentally windowed with a size of 5 (Appendix E). Thus, instead of extracting features from the 2D U/S frame, we extracted them from the volume in 256×192×5. In the U/S feature net, the input size was 256×192×5 and down/upsampled by 2 at each level. We did not downsample the data in the time dimension. The numbers of feature maps were 16, 32, 64, 128, and 256 in the encoder pathway and 256, 128, 64, 32, and 16 in the decoder pathway. The output layer was a convolutional layer with sigmoid activation and output of a U/S kidney feature map. Kidney feature maps of CT and U/S images were of the same dimension as the input. The windowed kidney feature images were used to construct the CT-US image pair.

### Registration Network

We present kidney registration (Fig. 1) in five aspects: image pair preprocessing, network structure, loss function, training data generation and learning strategy.

**Preprocessing:** All CT images were converted to RAI orientation and isotopically sampled in 0.8 mm × 0.8 mm × 0.8 mm. CT scans were automatically cropped to 128×224×288 around the centroid. A U/S image is resampled to 0.8 mm × 0.8 mm and cropped to 224×288. Windowed U/S images were centred in a 128×224×288 volume with zero padding. Within one U/S window, the middle was the registration target, and the others contributed to motion regularisation. U/S images were stacked along the time axis, the same as the R-L axis in image space. CT volume was aligned with U/S by matching the kidney centroids from feature maps. Because U/S scanning of the kidney was subject to the constraints of the ribs and spines of patients, the variability of the initial position of 3DCT-2DUS pairs can be large. To reduce variability, we uniformly aligned the kidney on CT in the inferior-superior axis and then aligned kidney on U/S with the centroid. The inputs to the registration network were CT and U/S window images of 128×224×288. CT was the moving image, and U/S was the fixed image.

**Network Structure:** The architecture is shown in Figure 1. The CT and U/S volumes were concatenated before going to the convolution layer. The numbers of feature maps were 8, 16, and 16 in the encoder pathway and 16, 16, 16, 16, 8, and 8 in the decoder pathway. Upsampling was performed using Upsampling3D, and downsampling was performed with stride convolution. Transformation regression used the affine block. At each decoder level, the activation of the output (dense) layer was tanh for transformation parameters. The spatial transformer network (STN) layer[38] processed rigid transformation because the kidney seldom underwent deformation during free respiration. The optimiser was Adam. The learning rate was 3e-4. Given that the window size of U/S images was $N_w$, the output transformation was a set of $6N_w$ rigid transformation parameters, $3 N_w$ for rotation and $3 N_w$ for translation. $N_w$ was experimentally set to 5 in this work (Appendix G). Each layer in the decoder pathway output a set of transformation parameters, and their weighted sum composed the final transformation. For rotation, it was an average of rotation parameters. For translation, it was a weighted sum of the translation parameters, and the weights were {8, 4, 2, 1}/4 = {2, 1, 0.5, 0.25}. Rotation was scale invariance and was averagely weighted. Translation was inversely proportional to the image resolution and was proportionally weighted. Hierarchical activation improved prediction and decreased the training time. The number of trainable parameters was approximately 282M.

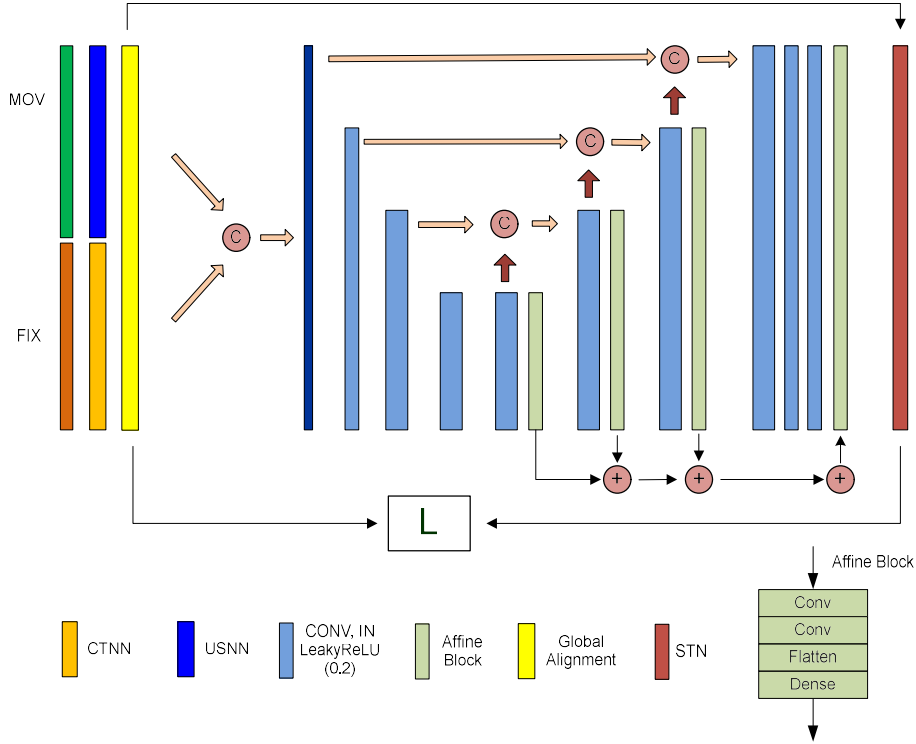

**Figure 1**. The 3D-2D registration network structure.

**Loss Function:** The feature map describes the overall kidney, and the MIND feature describes image details. They can complement each other to measure alignment accurately. As registration occurs during breathing, the CT cutting planes obtained should be able to stack together as smooth time sequences. Given $I_{fix}$ fixed image, $I_{mov}$ moving image, $M_{fix}$ fixed feature map, $M_{mov}$ moving feature map, and $\mathcal{D}$ transformation parameters, the feature-image-motion (FIM) loss was defined (Eq. (1)) by three measures on the feature map, original image, and transformation:

$$\mathcal{L}(I_{fix}, I_{mov}, M_{fix}, M_{mov}, \mathcal{D}) = \mathcal{L}_f(M_{fix}, \mathcal{D} \circ M_{mov}) + \lambda_1 \mathcal{L}_i(I_{fix}, \mathcal{D} \circ (I_{mov} \cdot M_{mov})) + \lambda_2 \mathcal{L}_d(\mathcal{D}) \quad (1)$$

$$\mathcal{L}_f(M_{fix}, \mathcal{D} \circ M_{mov}) = -\frac{1}{N_w} \sum_{i \in N_w} DICE_i(M_{fix}, \mathcal{D}_i \circ M_{mov}) \quad (2)$$

$$\mathcal{L}_i(I_{fix}, \mathcal{D} \circ (I_{mov} \cdot M_{mov})) = \frac{1}{|N|} \sum \left| MIND(I_{fix}) - MIND(\mathcal{D} \circ (I_{mov} \cdot M_{mov})) \right| \quad (3)$$

$$\mathcal{L}_d(\mathcal{D}) = 0.01 * \frac{\|\mathcal{D} - I_{4 \times 4}\|}{N_w} + 0.99 * grad\mathcal{D}, \text{ and } grad\mathcal{D} = \frac{1}{N_w - 2} \sum_{i-1,i,i+1 \in N_w} \|D_{i+1} + D_{i-1} - 2D_i\| \quad (4)$$

The feature loss was the negative Dice coefficient of the fixed and warped kidney feature map (Eq. (2)), where $DICE(x, y) = \frac{1}{m} \sum \left( \frac{2x \odot y}{x \oplus y} \right)$, $\odot$ and $\oplus$ are elementwise operations: multiplication and addition, and $m$ is the number of elements. DICE was calculated between the middle slice in R-L axis of warped CT volume and corresponding slice in fixed U/S volume. The feature loss was an average DICE within time windows. The image loss was the MIND feature difference between the fixed and warped original image (Eq. (3))[39]. A MIND feature was calculated by a Gaussian function of the mean squared difference between a central patch of the image and one of its six neighbouring patches. The neighbourhood was in 3D and the patch size was 3×3×3. MIND features were extracted from CT volume, and from U/S volume to capture 3D local image information as the kidney moved. U/S volume had 5 frames, and the feature of its middle frame was valid and used to calculate the image loss. The image loss was the mean absolute difference between the MIND features of middle slice in R-L axis in the warped CT volume and that in the fixed U/S volume to compare two images with local details. N was the set of displacement vectors. Here, the CT images needs to be masked before calculating MIND because CT images display extra anatomic structures in addition to the kidney. The transformation loss (Eq. (4)) was a weighted sum of the L2-norm divided by the number of parameters and the average transformation difference (Eq. (4)). $\lambda_1$, $\lambda_2$ were empirically set as 0.01 and 0.001. Here, we assumed that MIND is locally continuous. $\lambda_1$ was set to 0.01 to ensure that feature loss dominates weight updates initially, and when approaching the global optimum, image loss took effect to fine tune the weights. The transformation loss regularised the main transformation to move CT to its optimal position aligned with U/S images. Basically, feature loss, $\mathcal{L}_f$, was calculated on kidney feature maps. Image loss, $\mathcal{L}_i$, was computed between U/S images and masked CT images, which were implemented by elementwise multiplication of the warped CT images and the warped CT kidney feature map. Loss was calculated on kidney regions, thus eliminating influence from the difference between the field of views in CT and U/S images.

**Training Data Generation:** The transformation was in 6-dimensional space, while the data size was relatively small. Network training was prone to overfitting. Training dataset generation was necessary and was employed in projective 3D-2D image registration[25, 40]. Dense sampling was commonly used. Unlike projective registration, the transformation parameters or their distribution for sliced 3D-2D registration were uncontrollable and unavailable. We needed to find it out. First, we verified reference planes with clinicians (Appendix B). The transformations to obtain those verified alignments were modelled, parameter-by-parameter, in 2-sigma Gaussian distributions, where we randomly sampled $N_t$ transformations individually (Appendix C). The $N_t$ inverse transformations transform the optimal alignment back to $N_t$ different initial kidney positions. We generated $N_t$ training CT-US pairs by applying $N_t$ inverse transformations on a reference CT volume (Appendix D). Our training data generation scheme helped to obtain realistic training datasets to overcome overfitting. In the future, if sufficient clinical data are available, data generation can be neglected.

**Learning Strategy:** Based on the generated training datasets, unsupervised learning was employed to pretrain a registration model. Our target U/S images are respiration sequences including images of periodic breathing cycles. It was possible to make use of the onsite patient-specific dataset to further improve the model performance. We proposed a one-cycle transfer learning strategy, which refined the pretrained model using the first respiration cycle data via transfer learning with two epochs and inferred the transformations subsequently. On-site patient-specific training without pretrained model was infeasible because the convergence time was too long to accept, and a long operation preparation time was impractical in clinical applications. We proposed using transfer learning to refine the model to save time and improve performance (Appendix F).

**Evaluation Metrics:** The Hausdorff distance (HD) and mean contour distance (MCD) between the outlines of the kidney on CT and U/S images were used to evaluate if the CT-US pair was well aligned, and calculated as

$$D_{Hausdorff}(U,C) = \max\left\{\max_{c \in C}\left\{\min_{u \in U}\{d(c,u)\}\right\}, \max_{u \in U}\left\{\min_{c \in C}\{d(u,c)\}\right\}\right\}, D_{mean} = \operatorname*{mean}_{u \in U, c \in C}\{d(u,c)\}.$$

$d(u,c)$ is absolute value on the distance map, and $c$ and $u$ are contour points on the CT and U/S images, respectively. HD and MCD were in millimetres. The CT to U/S (CT-US) distance was calculated between kidney boundaries on CT and U/S images. The CT to CT (CT-CT) distance between kidney contours on the resulting CT plane and reference CT plane.

## Experimental Data

### Clinical Datasets

The datasets consisted of public KiTS19[41] datasets and in-house datasets. Public datasets consisted of 210 corticomedullary phase (CMP) CT images. The in-house datasets were collected from Fifth Affiliated Hospital Guangzhou Medical University and approved by the Institutional Review Board on August 28, 2020, with protocol number L2020-24. The datasets were consecutively studied from January to May 2021, consisting of 132 U/S image sequences (more than 30K images) from 31 patients, 39 multiple phase CT images from 24 patients, and 25 pairs of CT volumes and U/S sequences from 25 patients. All images were de-identified. The CT scans were acquired on one 64-slice scanner (GE OPTIMA CT600 CT scanner), using a standard four-phase contrast-enhanced imaging protocol with a slice thickness of 0.6–5.0 mm, a matrix of 512×512 pixels and an inplane resolution of 0.625–0.916 mm. CMP scanning was performed with 180 HU detected at the region of interest within the abdominal aorta. The nephrographic phase (NP) was performed 28 s post contrast, and the excretory phase (EP) was performed 10–30 min post contrast. The U/S datasets were acquired on a GE Versana Active™ ultrasound system, with a matrix of 1132×852 pixels and an inplane resolution of 0.22–0.29 mm. The U/S sequences were scanned at 17-22 frames per second and had 58±14 U/S images per respiration cycle.

### Generated Training Datasets

**Reference planes:** For each U/S frame, a manually selected reference CT cutting plane with the overlap of kidney boundaries was displayed side-by-side to four experienced clinicians to unanimously verify if they were the same cutting plane of the kidney. The verified planes constructed our basic reference set, from which we extended the training set.

There were 22 out of 25 pairs of CT volume and U/S sequences verified by clinicians that their 3D cutting plane in CTs was the same as that in U/S images.

**Transformation Parameter Estimation:** Six parameters in transformations that resulted in the reference CT planes from initial positions were modelled in 2-sigma Gaussian distributions. The clinical datasets were split in a ratio of 7:1:2 for training, validation and testing sets according to patients. They were normalised to pixels having a zero mean value and one variance value. Fivefold cross validation was used to evaluate the model's performance. The patient's data was split into five groups. One group was testing dataset and the remaining groups were train/validation dataset. The train/validation dataset was randomly split. The performance of the model was averaged over five runs. Only training datasets were generated N times.

## Experiments and Results

### Evaluation on Feature Network

Our network was implemented on TensorFlow, and training was performed on a workstation with a dual Nvidia Quadro RTX 5000 of 16 GB and CPU memory of 256 GB. The learning rate was $1 \times 10^{-4}$. The model was evaluated by

$$DICE = \frac{2|Y \cap Y^*|}{|Y| + |Y^*|}, \quad Sensitivity = \frac{|Y \cap Y^*|}{|Y|}, \quad Specificity = \frac{|Y \cap Y^*|}{|Y^*|},$$ where $Y^*$ is prediction and $Y$ is ground truth.

The model resulted in an average Dice coefficient of 96.88% on kidney segmentation in the plain phase CT images (Table 1). The network resulted in an average Dice coefficient of 96.39% on U/S images.

|      | DICE   | Sensitivity | Specificity |
|------|--------|-------------|-------------|
| CT   | 0.9688 | 0.9711      | 0.9667      |
| U/S  | 0.9639 | 0.9736      | 0.9560      |

**Table 1**: Feature network performance on kidney segmentation in CT and U/S images.

### Evaluation on Registration Network

The generated datasets were only used for training. The more datasets generated, the smaller the MCD achieved. Here, we generated approximately 12,000 training data pairs, 10 times of clinic datasets, to pretrain the registration network. We conducted an ablation study on the method (Table 2). All the network components contributed to performance improvement. The hierarchical transformation regression at decoder layers contributed more than the MIND loss. One-cycle transfer learning contributed the most. The uncertainty estimation was in Appendix H.

| Methods | Metric (mm) | CT-CT | CT-US |
|---------|-------------|-------|-------|
| Ours | HD | 3.97±1.37 | 3.80±0.87 |
|  | MCD | **1.15±0.57** | **0.94±0.18** |
| Without MIND Loss | HD | 4.43±1.40 | 3.74±0.90 |
|  | MCD | 1.38±0.56 | 0.98±0.23 |
| Without Decoder Layers | HD | 7.66±2.95 | 4.14±1.15 |
|  | MCD | 2.45±1.10 | 1.02±0.20 |
| Without Hierarchical Transformation regression | HD | 8.15±3.94 | 4.16±0.97 |
|  | MCD | 2.71±1.48 | 1.05±0.21 |
| Without One Cycle Transfer Learning | HD | 11.59±1.57 | 6.71±1.06 |
|  | MCD | 4.09±0.65 | 2.03±0.29 |

**Table 2**: CT-US kidney registration performance in the ablation study with individual components removed.

### Result Comparison

As our deep learning-based pipeline is the first for 3DCT-2DUS kidney registration, we can only compare our method with SOTAs [14, 32, 33] in terms of the registration module (Table 3). We replaced our encoder-decoder hierarchical registration subnet with the encoder structures or transformer-based registration module. That is, the input to SOTA was CT-US feature pairs after global alignment. VoxelMorph, ConvNet-affine[14], VTN-affine[32], and C2FViT[33] were trained on the general training datasets to converge. VoxelMorph uses the same affine block as in our registration network to obtain rigid transformation parameters. ConvNet-affine and VTN-affine were implemented based on their papers while rigid transformation was employed.

Ours, with two-step training, was superior to all SOTAs when measured by HD and MCD (Table 3). The 2D CT-US distances were smaller than the 3D CT-CT distances because the distance in one dimension was overlooked. With a one-step training strategy, our pretrained model performed better than VoxelMorph with a smaller 3D CT-CT distance and a larger 2D CT-US distance. This result indicated that hierarchical structures prevented convergence to a local optimum. In addition, it was found that our pretrained model with hierarchical structures converged at approximately 20-50 epochs during training, much faster than VoxelMorph, converging at approximately 200-300 epochs. In addition, the transformer-based method, C2FViT, performed better than CNN-based methods with a one-step training strategy.

| Methods | Metric (mm) | One-step learning | | Two-step learning | |
|---|---|---|---|---|---|
| | | CT-CT | CT-US | CT-CT | CT-US |
| Ours | HD | 11.59±1.57 | 6.71±1.06 | 3.97±1.37 | 3.80±0.87 |
| | MCD | 4.09±0.65 | 2.03±0.29 | **1.15±0.57** | **0.94±0.18** |
| VoxelMorph | HD | 12.21±1.87 | 5.99±0.93 | 8.15±3.94 | 4.16±0.97 |
| | MCD | 4.38±0.72 | 1.82±0.28 | 2.71±1.48 | 1.05±0.21 |
| C2FViT | HD | 10.36±1.77 | 6.42±0.84 | 9.93±1.72 | 5.64±0.90 |
| | MCD | **3.74±0.91** | **1.93±0.50** | 3.48±0.78 | 1.45±0.42 |
| VTN-Affine | HD | 12.18±1.86 | 5.58±1.01 | 7.66±2.95 | 4.14±1.15 |
| | MCD | 4.21±0.71 | 1.45±0.35 | 2.45±1.10 | 1.02±0.20 |
| ConvNet-Affine | HD | 11.58±2.43 | 6.72±1.34 | 4.56±1.55 | 3.89±0.91 |
| | MCD | 4.36±1.11 | 2.07±0.51 | 1.46±0.62 | 1.00±0.23 |

**Table 3**: Registration performance comparison with SOTAs using a one-step learning strategy (without one-cycle transfer training applied) and using a two-step learning strategy (with one-cycle transfer learning applied).

We compared our method with SOTAs using a two-step training strategy, all with a one-cycle transfer learning strategy applied (Table 3). Ours learned most from transfer learning to improve performance. ConvNet, a Siamese encoder structure, was second only to ours. C2FViT learned the least. The CNN-based method performed better than the transformer-based method with one-cycle transfer learning for two epochs. Our method performed best. Example results were in Appendix I.

We divided the CT-US image pairs into two groups sorted by transformations. Group A comprised datasets of small transformations, rotations of 10.37±2.24 degrees and translation of 3.69±0.95 mm. Group B comprised datasets of large transformations, rotations of 24.72±2.28 degrees and translation of 5.04±1.20 mm. Rotation and translation were separately calculated as the L2-norm of components in the x, y, and z directions. All deep learning-based methods used a two-step learning strategy. The performance of the two groups was measured (Table 4). Our method performed best on both Groups A and B. It was robust to large transformations.

| Methods | Metric (mm) | Group A | | Group B | |
|---|---|---|---|---|---|
| | | CT-CT | CT-US | CT-CT | CT-US |
| Ours | HD | 3.58±1.37 | 3.85±0.32 | 4.36±1.46 | 3.75±0.83 |
| | MCD | **1.02±0.59** | **0.82±0.04** | **1.28±0.53** | **1.10±0.20** |
| Voxel-Morph | HD | 7.97±0.31 | 4.20±0.34 | 8.34±6.75 | 4.12±0.94 |
| | MCD | 2.63±0.02 | 0.95±0.04 | 2.79±2.49 | 1.15±0.23 |
| C2FViT | HD | 8.01±0.09 | 5.82±0.88 | 11.85±1.79 | 5.47±0.09 |
| | MCD | 3.38±0.65 | 1.37±0.23 | 3.57±0.58 | 1.53±0.10 |
| VTN-Affine | HD | 5.84±0.41 | 4.41±0.47 | 9.48±5.21 | 3.86±1.09 |
| | MCD | 1.83±0.13 | 0.95±0.02 | 3.08±1.94 | 1.09±0.20 |
| ConvNet-Affine | HD | 4.32±1.06 | 3.72±0.01 | 4.80±1.66 | 4.06±1.04 |
| | MCD | 1.28±0.43 | 0.83±0.04 | 1.64±0.72 | 1.16±0.28 |

**Table 4**: Registration performance on datasets with small and large transformations.

Example testing sequences of CT and U/S images were displayed in RAI orientation (Fig. 2). U/S frames and resulting CT planes in sagittal view were stacked along the R-L orientation. The axial and coronal views provided dynamic information and the R-L axis represented time. The coronal view displayed the up and down motion of the kidney during respiration, and the axial view displayed the back-and-forth motion of the kidney.

Finding corresponding landmarks between CT and U/S was difficult. We identified a special case that had a small lesion visible on both the U/S images and CT images during breathing (Fig. 3). The lesion was approximately 7 mm in diameter on the left kidney. It was interesting to know the tumour distance after registration for this special case. If we assumed that the centre of the lesion observed on U/S scan was also its centroid on CT volume, the tumour centre distance was 3.39 mm after registration. 4.38 mm, 6.04 mm, 4.16 mm, and 3.91 mm resulted from VoxelMorph, C2FViT, VTN-affine, and ConvNet-affine, respectively. The distance was large because the tumour was near the kidney surface.

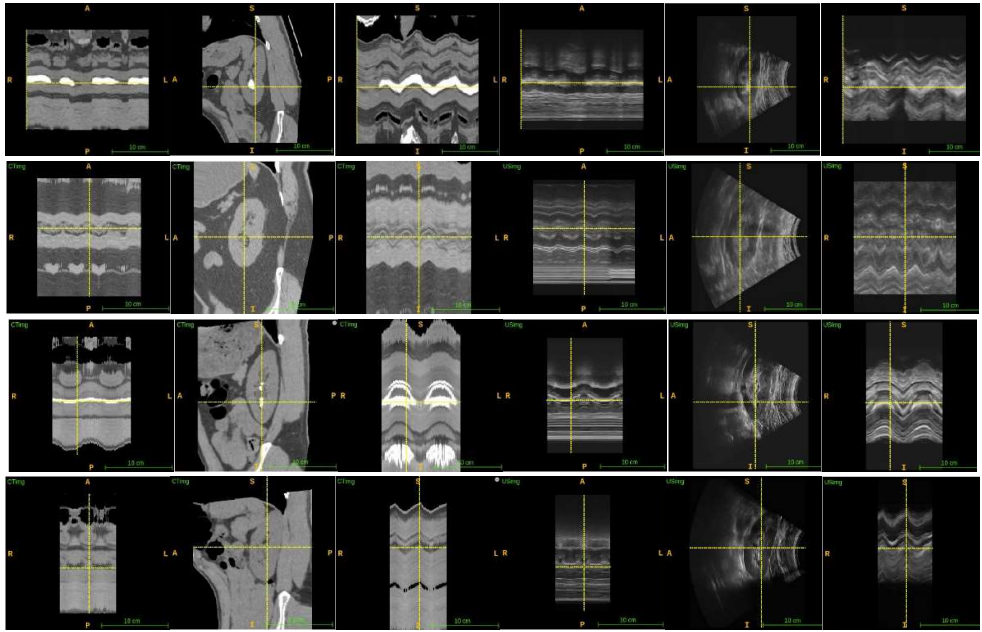

**Figure 2**. Example registration results, from left to right, the images are CT axial view, sagittal view, coronal view, U/S axial view, U/S sagittal view, U/S coronal view. Rows 1 and 3 visualized internal alignment of renal stone (crosshairs in the sagittal view). Rows 2 and 4 visualized slice alignment on inspiration and expiration (crosshairs in the coronal view).

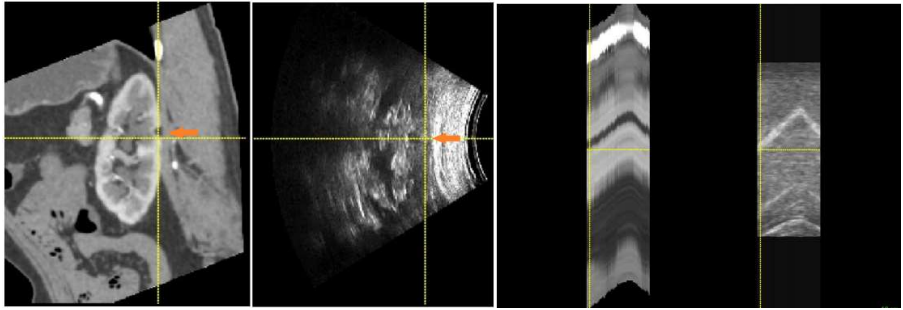

**Figure 3**. Example image pair to visualize target registration error (TRE) on a small lesion in CT image and U/S image (left), and small lesion's motion in respiration (right).

**Performance Comparison With 2D/3D Registration Methods**

After unsupervised affine registrations were compared, in this section, we presented the performances of existing 2D/3D registration methods in Table 5. SOTA methods addressed deep 2D/3D registration using supervised CNN regression[6, 25, 27, 30], supervised reinforcement learning[28, 29], or supervised CNN segmentation cum conventional plane regression[31]. They were all supervised learning. Our method used unsupervised end-to-end convolutional neural network which output both transformation parameters and resulting cutting plane, was trained without the ground truth transformations, and achieved comparable performance with the supervised methods.

## Discussion

It is challenging to obtain perfect spatial correspondence for CT and U/S images. We measured the distance on the reference CT-US pairs; the average HD was 3.57±1.11 mm, and the MCD was 0.79±0.22 mm, which was approximately one pixel in size. The nonzero distance may be due to imperfect contour extraction, rib occlusion, or different patient postures.

The pretrained model benefits using training data generation, because the transformation parameters are in high dimensional space while the data size is small. However, only increasing the generated datasets makes it difficult to improve the model performance further. We proposed using transfer learning to achieve this goal.

| 2D/3D Methods | Anatomy | Modality | Deep learning | Resolution | Mean Error |
|---|---|---|---|---|---|
| Miao (2016) [6] | Phantom | X-rays / DDR | Supervised CNN regression | 0.223 mm | 0.282 mm |
| Foote (2019) [25] | Lung | CT | Supervised CNN regression | 0.388 mm | 1.22 mm |
| Salehi (2018) [27] | Fetal brain | MR | Supervised CNN regression | 0.8 mm | 13.08 degrees (rotation) 2 mm (translation) |
| Guo (2021) [30] | Prostate | TRUS | Supervised CNN regression | - | 2.73 mm |
| Liao (2017) [28] | Spine | CBCT/CT | Supervised reinforcement learning | - | 3.8 mm |
| Krebs (2017) [29] | Prostate | MR | Supervised reinforcement learning | 2 mm | 7.7 mm |
| Wei (2021) [31] | Liver | CT/US | Supervised CNN segmentation + post plane regression | 4 mm | 11.6 degrees (rotation) 4.7 mm (translation) |
| Ours | Kidney | CT/US | **Unsupervised end-to-end CNN** | 0.8 mm | **1.15 mm** |

**Table 5**: Registration performance of existing 2D/3D deep learning methods.

Transfer learning may need extra preparation/training time before application. With a pretrained model, the preparation can be shortened significantly. For example, if the registration model was trained from scratch using one-cycle learning, it took approximately 46 mins to converge, while only 2-3 mins were needed if the model was trained using transfer learning. Thus, a good pretrained model was essential to practical applications.

Even though optimal alignments verified with clinicians were available, we did not change our unsupervised learning to supervised learning. First, the data size was quite small, and much effort was required to obtain more data. It was not desirable to limit our model from processing versatile pairing data when available. The increased training data generated a regularisation effect, which benefited the cost function optimisation, and reduced of overfitting and model generalization. We believe that in the future, it will be possible to collect more paired datasets to overcome overfitting resulting from the dataset's limitation.

There was conventional method proposed by Wein[42] on the 3DCT-3DUS kidney registration. The U/S images were acquired using a tracked probe during breath-hold on inspiration. Optimisation was performed by exhaustive search of translation space. In this work, we aimed at a deep learning-based 3DCT-2DUS kidney registration during breathing, which performed deep model inference.

## Conclusions

To the best of our knowledge, this paper presented the first deep learning pipeline for sliced 3DCT-2DUS kidney registration. All difficulties in kidney registration during free breathing were addressed via novel network structures and training strategies. Comprehensive experiments showed that our proposed methodology performed well.

## Acknowledgement

This work is supported by ACCL/19-GAP035-R20H.

## Ethics Declarations

The in-house datasets were collecting from the fifth affiliated hospital Guangzhou Medical University, approved by the Institutional Review Board on Aug 28, 2020 with protocol number of L2020-24.

# Appendix

## A: Feature Network

The ULBNet is a 5-level U-Net with residual block replacing the original convolutional layer (Fig.1).

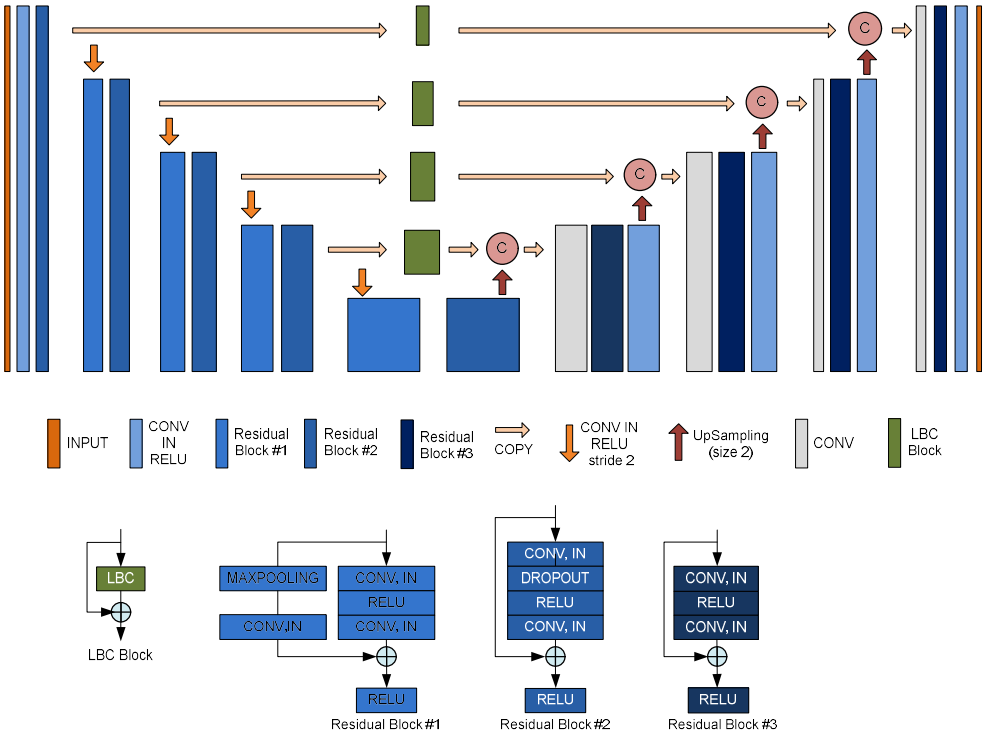

**Figure 1**. The 3D ULBNet network architecture.

## B: Reference Plane

For each U/S frame, a manually selected reference CT cutting plane with the overlap of kidney boundaries was displayed side-by-side to four experienced clinicians to unanimously verify if they were the same cutting plane of the kidney. The verified planes constructed our basic reference set, from which we extended the training set. There were 22 out of 25 pairs of CT volume and U/S sequences verified by clinicians that their 3D cutting plane in CTs was the same as that in U/S images (Fig. 2).

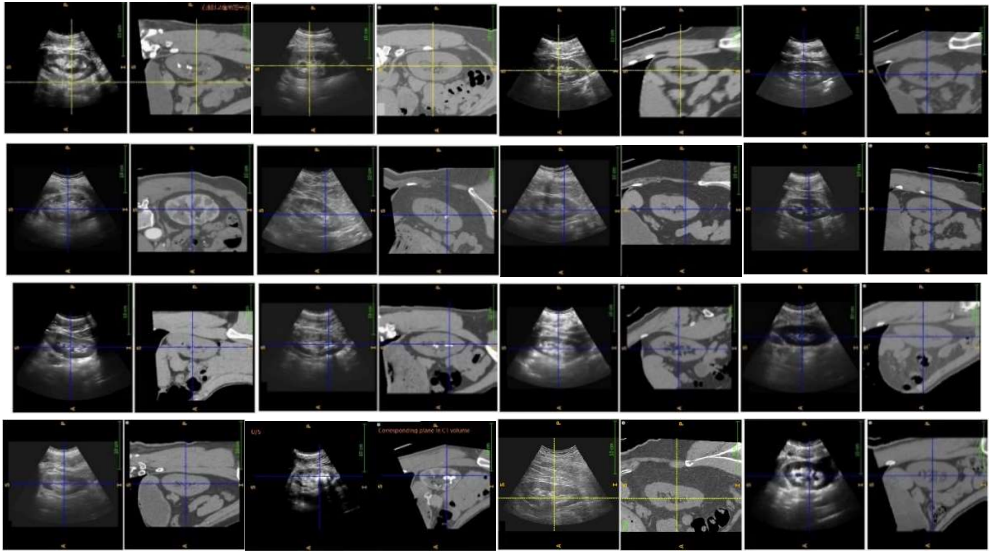

**Figure 2**. Examples of the verified reference planes.

## C: Transformation Parameter Estimation

Six parameters in transformations that resulted in the reference CT planes from initial positions (Fig. 3) were modelled in 2-sigma Gaussian distributions.

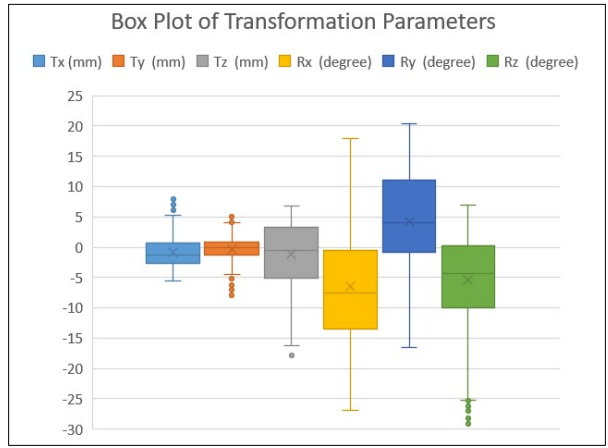

**Figure 3**. Distribution of transformation parameters (T: translation, R: Euler angle)

## D: Generated Datasets

The generated datasets were only used for training. The more datasets generated, the smaller the MCD achieved (Fig.4). A local minimum occurred at 6000 pairs. The global minimum occurred approximately 12,000 pairs, where the 3D CT-CT distance was close to 2D CT-US distance. Thus, we generated approximately 12,000 training data pairs, 10 times of clinic datasets, to pretrain the registration network.

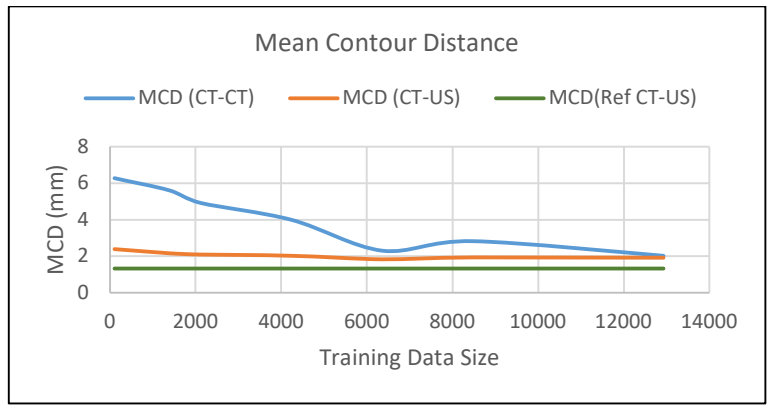

**Figure 4**. The average MCD decreases when the generated training datasets increase.

## E: U/S Feature Network Performance on Window Size

We tested feature extraction on U/S image sequences with different window sizes (Table 1). The best performance was obtained at a size of 5.

| Window size | DICE | Sensitivity | Specificity |
|---|---|---|---|
| 1 | 0.9391 | 0.9492 | 0.9455 |
| 3 | 0.9543 | 0.9714 | 0.9407 |
| 5 | **<u>0.9639</u>** | **<u>0.9736</u>** | 0.9560 |
| 7 | 0.9604 | 0.9675 | **<u>0.9561</u>** |
| 9 | 0.9590 | 0.9656 | 0.9544 |

**Table 1**: Feature network performance on kidney segmentation in U/S images with different window sizes. The U/S images were of dimension 256×192×WindowSize.

## F: Registration Performance on Transfer Learning Strategy

The model performance on four test sequences from the second cycle was evaluated. Overall, one-cycle training performed best, as observed in Table 2. Moreover, we found that transfer learning helped to achieve a lower loss than the model trained from scratch, as shown in Table3.

| | Metric (mm) | | Learning strategy | | | |
|---|---|---|---|---|---|---|
| | | | Pretrained model | One-shot (5epoches) | Five-shot (5epoches) | One-cycle (2epoches) |
| Case 01 | CT-US | HD | 7.25±1.22 | 3.99±0.86 | 3.20±0.72 | 3.01±0.73 |
| | | MCD | 2.53±0.30 | 1.07±0.32 | 0.85±0.18 | 0.75±0.17 |
| | CT-CT | HD | 5.78±1.50 | 6.12±1.66 | 6.00±1.70 | 6.75±2.42 |
| | | MCD | 2.03±0.56 | 2.06±0.56 | 1.80±0.52 | 2.17±0.95 |
| Case 02 | CT-US | HD | 8.32±1.21 | 4.91±0.77 | 4.35±1.22 | 4.16±0.65 |
| | | MCD | 2.11±0.46 | 1.61±0.70 | 0.95±0.20 | 0.85±0.13 |
| | CT-CT | HD | 19.18±2.66 | 4.52±1.34 | 4.80±1.69 | 4.45±2.09 |
| | | MCD | 6.38±0.84 | 2.00±1.01 | 1.56±0.62 | 1.47±0.91 |
| Case 03 | CT-US | HD | 5.39±0.72 | 4.89±0.90 | 3.61±0.99 | 3.54±1.11 |
| | | MCD | 1.50±0.19 | 1.36±0.38 | 0.95±0.27 | 0.79±0.19 |
| | CT-CT | HD | 16.52±1.21 | 11.92±2.32 | 4.68±1.44 | 2.72±0.64 |
| | | MCD | 5.70±0.39 | 3.93±0.82 | 1.64±0.84 | 0.58±0.26 |
| Case 04 | CT-US | HD | 5.89±1.11 | 5.29±1.16 | 4.45±1.10 | 4.52±0.99 |
| | | MCD | 1.99±0.23 | 1.75±0.28 | 1.43±0.23 | 1.40±0.23 |
| | CT-CT | HD | 4.90±0.92 | 4.68±0.85 | 2.94±0.79 | 1.98±0.34 |
| | | MCD | 2.27±0.84 | 2.00±0.86 | 0.89±0.45 | 0.39±0.19 |
| Average | CT-US | HD | 6.71±1.06 | 4.77±0.92 | 3.90±1.00 | 3.80±0.87 |
| | | MCD | 2.03±0.29 | 1.44±0.42 | 1.04±0.22 | **<u>0.94±0.18</u>** |
| | CT-CT | HD | 11.59±1.57 | 6.81±1.54 | 4.63±1.40 | 3.97±1.37 |
| | | MCD | 4.09±0.65 | 2.49±0.81 | 1.47±0.60 | **<u>1.15±0.57</u>** |

The CT-US distance is calculated between kidneys boundaries on CT and U/S. The CT-CT distance between kidney contours on resulted CT plane and reference CT plane.

**Table 2**: 3DCT-2DUS kidney registration performance on four learning strategies.

| | Loss (Epoch = 2) | | Loss$_{min}$ | |
|---|---|---|---|---|
| | TFL | TFS | TFL | TFS |
| Case 01 | -0.9744 | -0.9681 | -0.9773 | -0.9762 |
| Case 02 | -0.9642 | -0.9508 | -0.9689 | -0.9673 |
| Case 03 | -0.9624 | -0.9507 | -0.9706 | -0.9699 |
| Case 04 | -0.9610 | -0.9559 | -0.9692 | -0.9683 |
| Average | **<u>-0.9655</u>** | -0.9563 | **<u>-0.9715</u>** | -0.9704 |

TFL: training model via transfer learning, TFS: training model from scratch, Loss$_{min}$: the minimal loss obtained within 100 epochs.

**Table 3**: Feature-image-motion loss obtained from two learning strategies: TFL and TFS.

**G: Registration Network Performance on Window Size**

We tested registration on motion regularisation with window sizes of 1, 3, and 5 (Table 4), and a size of 1 resulted in best accuracy, while a size of 5 resulted in visually smoother CT cutting plane sequences. It was reasonable that the large window size gave a smooth regularisation at the expense of accuracy. The size of 1 approached the global minimum by free transition and rotation, while the size of 5 approached the optimal transformation by transition on a larger scale due to motion regularisation. Since smoothness was also important, we selected 5 to achieve overall optimality.

| WindowSize | Metric (mm) | CT-CT | CT-US |
|---|---|---|---|
| 1 | HD | 2.91±1.11 | 3.73±0.95 |
| | MCD | **0.97±0.44** | 0.95±0.21 |
| 3 | HD | 3.39±1.38 | 3.82±0.96 |
| | MCD | 0.99±0.58 | 0.98±0.25 |
| 5 | HD | 3.97±1.37 | 3.80±0.87 |
| | MCD | 1.15±0.57 | **0.94±0.18** |

**Table 4**: CT-US registration performance on the number of consecutive frames (WindowSize) used for motion regularisation.

**H: Uncertainty Estimation**

Due to the unavailability of ground truth transformations for image registration, an uncertainty estimate would help gauge system reliability. Registration uncertainty was estimated by randomly removing 1~10% training datasets and estimating the standard deviation of the performance. The experiment was repeated (n=10). The uncertainty in MCD was approximately 1 mm for the CT-CT distance and 0.22 mm for the CT-US distance (Table 5).

| | CT-US | | CT-CT | |
|---|---|---|---|---|
| | HD | MCD | HD | MCD |
| Standard deviation (mm) | 1.62 | 0.22 | 2.97 | 1.01 |

**Table 5**: Performance variance in ten repeated experiments on CT-US registration with random training datasets removed.

**I: Registration Results Examples**

We illustrated the registration results from the two-step training strategy by plotting the U/S image plane in CT volume (Fig. 5). The initial position (blue), the reference plane position (green), and the result plane position (red) can be observed in the coronal view. Our method resulted in the closest cutting plane to the reference plane compared to the other methods.

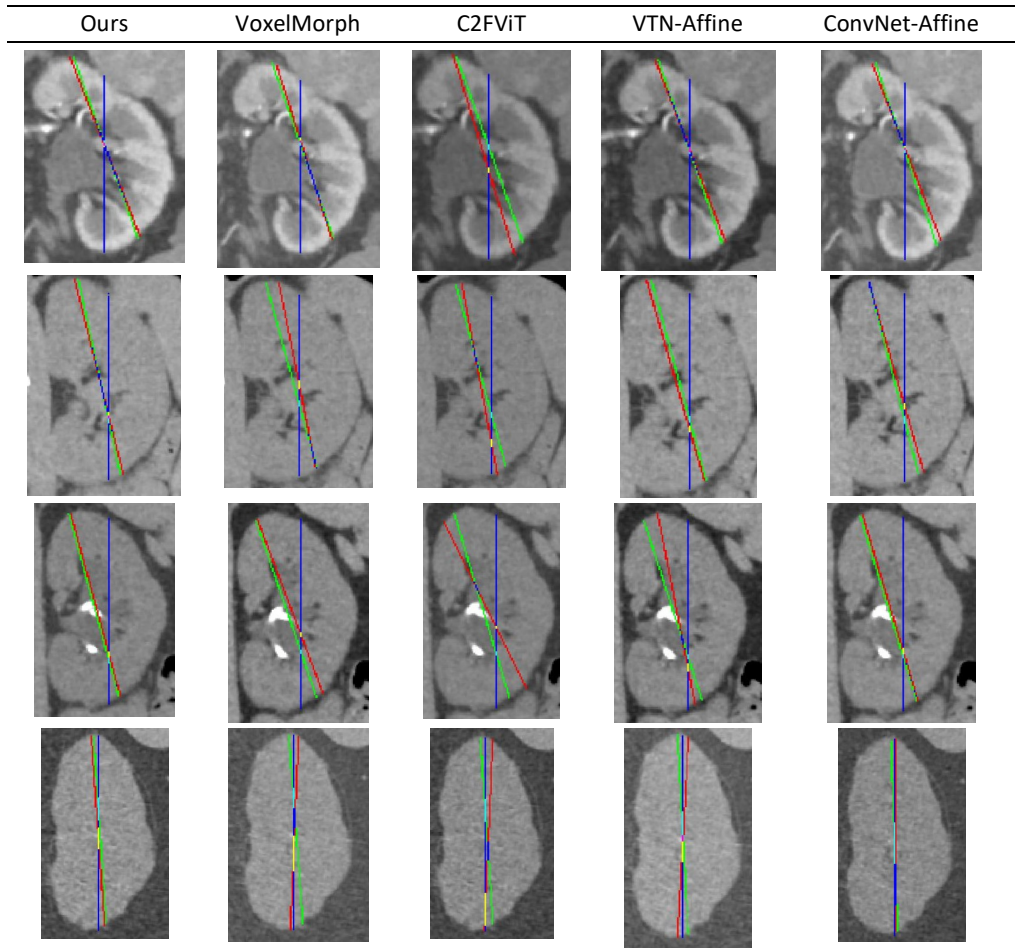

**Figure 5**. Example of the U/S plane in CT volume, displayed in coronal view. Blue: initial position, green, reference position, red: registration position.